\documentstyle[aps,preprint]{revtex}

\begin{document}

\newcommand{\ben}{\begin{equation}}
\newcommand{\een}{\end{equation}}
\newcommand{\bena}{\begin{eqnarray}}
\newcommand{\eena}{\end{eqnarray}}
\newcommand{\taubar}{\mbox{$\bar{\tau}$}}
\newcommand{\taubarp}{\mbox{$\bar{\tau}_p$}}
\newcommand{\omz}{\mbox{$\omega_{0}$}}
\newcommand{\xp}{\mbox{$x_{+}$}}
\newcommand{\xm}{\mbox{$x_{-}$}}
\newcommand{\tzt}{\mbox{$T^{(c)}_{0}$}}

\newcommand{\thbar}{\mbox{$\bar{\theta}$}}
\newcommand{\phibar}{\mbox{$\bar{\phi}$}}
\newcommand{\cth}{\mbox{$\vartheta$}}
\newcommand{\cphi}{\mbox{$\varphi$}}
\newcommand{\thz}{\mbox{$\vartheta_{0}$}}

\draft

\title{
Level splittings in exchange-biased spin tunneling
}

\author{
Gwang-Hee Kim\cite{e-kim}
}

\address{Department of Physics, Sejong University,
Seoul 143-747, Republic of Korea}
\date{Received \hspace*{10pc}}
\maketitle

\thispagestyle{empty}

\begin{abstract}
The level splittings in a dimer with the antiferromagnetic coupling
between two single-molecule magnets are calculated
perturbatively for arbitrary spin. It is found that
the exchange interaction between two single-molecule magnets
plays an important role in the level splitting.
The results are discussed in comparison with the recent
experiment.
\\
\end{abstract}

The quantum properties of single-molecule magnets have generated
considerable interest over the past decade in connection with 
macroscopic quantum phenomena\cite{gun}. High-spin molecules with
spin-10, $\rm Mn_{12}$ and $\rm Fe_8$ have been such good
candidates because all the clusters are identical with no dispersion
on the size of the clusters and the number of interaction spins, and the
spin ground state and the magnetic anisotropy are known with
great accuracy. These molecules display particularly interesting
phenomena such as  quantum resonant tunneling\cite{tho,san} and quantum phase
interference\cite{los}. Such phenomena have received much attention, both
theoretically and experimentally in view of macroscopic
realization of quantum tunneling, and also because of some
potential application to quantum computing\cite{leu}.
Many efforts have been made to understand their mechanisms
by considering a giant spin Hamiltonian with a 
single-molecule magnet\cite{tho,san,leu,gar}.
Most of the study have neglected exchange interactions
that depend on the distance and the non-magnetic atoms in the
exchange pathway. Recently, however, it has been reported that
a supramolecular single-molecule magnet dimer with antiferromagnetic
coupling exhibits quantum behavior different from that of the individual
single-molecule magnets\cite{wer}. This result implies that
exchange interaction between two single-molecule magnets can
have a large influence on the quantum properties of single-molecule magnets.
It is therefore important to understand the effect
of the exchange interaction on magnetization tunneling.

The issue of spin tunneling with the exchange interaction has been
raised by several groups\cite{bar}. In their studies exchange interaction is enhanced
to magnetic anisotropy for studying tunneling of the N\'eel  vector in antiferromagnetic
particles. Using the instanton technique based on spin coherent state path
integral, they calculated the tunneling rate of the N\'eel  vector in uniaxial
or biaxial antiferromagnetic particles. However,
the previous works applicable in the limit $S \gg 1$
have been confined to the spin tunneling of the ground state in an antiferromagnetic
particle having two collinear ferromagnetic sublattices.
In this paper, we will study magnetic tunneling in a system of identical, 
antiferromagnetically
coupled dimer. By employing a perturbative approach\cite{garJPA},
we obtain the level splitting of the states degenerate pairwise for arbitrary
spin in some typical cases and show that even weak exchange interaction plays
a crucial role in inducing spin tunneling.

The spin Hamiltonian of the dimer system can be written in the form
\bena
{\cal H}={\cal H}_1+{\cal H}_2 +J \hat{S}_1 \cdot \hat{S}_2,
\label{hamil}
\eena
where ${\cal H}_i$ ($i=1, 2$) is the Hamiltonian of each single-molecule
magnet which can be modeled as a giant spin of $S_i$.
The corresponding Hamiltonian is given by
\bena
{\cal H}_i =
-D \hat{S}^{2}_{zi}+{\cal H}^{\rm trans}_{i} -  H_z  \hat{S}_{zi},
\eena
where $D$ is the anisotropy constant and ${\cal H}^{\rm trans}_{i}$ 
includes the transverse anisotropy or field.
Also, ${\bf H}$ stands for $g \mu_B {\bf H}$ where
$g$ is the electronic $g$-factor and $\mu_B$ is the Bohr magneton.
Henceforth, we will usually drop the combination $g \mu_B$ for better
readability of the formula.
Since the dimer consists of
two single-molecule magnets with antiferromagnetic coupling,
we take $J>0$ much less than the anisotropy constant $D$.
The system has $(2S_1 +1)(2 S_2 +1)$ degenerate energy levels which in
the absence of the transverse terms of Eq. (\ref{hamil}) are
labeled by the spin projection $M_1$ and $M_2$ on the $z$-axis and
given by $E_{M_1,M_2}=-D (M^{2}_{1}+M^{2}_{2})+J M_1 M_2$.
It can be easily checked that for the longitudinal field
$H_z$ satisfying
\bena
H_z=
{D(M^{2}_{1}+M^{2}_{2}-{M^{\prime}_{1}}^{2}-{M^{\prime}_{2}}^{2} )
+J(M^{\prime}_{1}M^{\prime}_{2}-M_{1}M_{2})
\over M^{\prime}_{1}+M^{\prime}_{2}-M_{1}-M_{2}},
\label{longi-field}
\eena
the energy levels are degenerate:
\bena
E_{M^{\prime}_{1},M^{\prime}_{2}}=E_{M_1,M_2}.
\eena
Tunneling among the $(2S_1+1)(2S_2+1)$ energy states is allowed by the
transverse terms containing $\hat{S}_{xi}$ and $\hat{S}_{yi}$. In the case
of small transverse terms which is relevant for the dimer, the level
splittings can be calculated in a more direct and simple way
using the high-order perturbation theory.
In such cases, the level splitting of the degenerate level pair
$(M^{\prime}_{1},M^{\prime}_{2})$ and $(M_1, M_2)$ is represented
as the shortest chain of matrix elements and energy denomenators connecting
the states $|M^{\prime}_{1},M^{\prime}_{2}\rangle$ and $|M_1,M_2\rangle$
for the typical situations which will be considered.

Let us consider as model I the level splitting induced by the
transverse terms in the exchange interaction:
\bena
{\cal H}=-D \hat{S}^{2}_{z1} - D \hat{S}^{2}_{z2}+
J \hat{S}_1 \cdot \hat{S}_2.
\label{hamil1}
\eena
Noting that
$\hat{S}_1 \cdot \hat{S}_2 =\hat{S}_{1z}\hat{S}_{2z}+
{1 \over 2} (\hat{S}_{1+}\hat{S}_{2-}+\hat{S}_{1-}\hat{S}_{2+})$
and considering $\hat{S}_{1-}\hat{S}_{2+}$,
the level splitting of the degenerate pair
$(M^{\prime}_{1}, M^{\prime}_{2})$, $(M_1,  M_2)$
appears only in the chain of matrix elements with
connecting the states $| M^{\prime}_{1}+k, M^{\prime}_{2}-k \rangle$
and $| M^{\prime}_{1}+k+1, M^{\prime}_{2}-k-1 \rangle$ where
$M^{\prime}_{1}=-M_1$, $M^{\prime}_{2}=M_1 >0$, 
$M_2=-M_1$, and
$k$  is an integer with $0 \leq k \leq M_1 -1-M^{\prime}_{1}$.
It corresponds to the level splitting of
the degenerate pair $(-M_1, M_1) \rightarrow (M_1, -M_1)$.
In this case the magnetic field does not contribute
to the level splitting and thereby
the longitudianl field (\ref{longi-field}) is not taken into consideration.
Then, the level splitting of the degenerate pair
becomes
\bena
\Delta E_{M^{\prime}_{1} M^{\prime}_{2}, M_1  M_2}&=&
2V_{M^{\prime}_{1} M^{\prime}_{2}, M^{\prime}_{1}+1,M^{\prime}_{2}-1}
{1 \over E_{M^{\prime}_{1}+1, M^{\prime}_{2}-1}-
E_{M^{\prime}_{1}M^{\prime}_{2} }} 
V_{M^{\prime}_{1}+1, M^{\prime}_{2}-1, M^{\prime}_{1}+2,M^{\prime}_{2}-2}
\nonumber \\
&&\times 
{1 \over E_{M^{\prime}_{1}+2, M^{\prime}_{2}-2}-
E_{M^{\prime}_{1}M^{\prime}_{2} }}...
V_{M_{1}-1, M_{2}+1, M_{1}M_{2}},
\label{pro}
\eena
where
\bena
V_{M^{\prime}_{1} M^{\prime}_{2}, M^{\prime}_{1}+1,M^{\prime}_{2}-1}&=&
\langle M^{\prime}_{1} M^{\prime}_{2}| 
{J \over 2} (\hat{S}_{1+}\hat{S}_{2-}+\hat{S}_{1-}\hat{S}_{2+})
|M^{\prime}_{1}+1,M^{\prime}_{2}-1\rangle \nonumber \\
&=& {J \over 2} l_{M^{\prime}_{1}+1,M^{\prime}_{2}-1},
\eena
$l_{M^{\prime}_{1}+1,M^{\prime}_{2}-1}=\sqrt{(S_1 +M^{\prime}_{1}+1)
(S_1-M^{\prime}_{1})(S_2 -M^{\prime}_{2}+1)(S_2+M^{\prime}_{2})}$
are the matrix elements of the operator 
$\hat{S}_{1-}\hat{S}_{2+}$, and
$E_{M^{\prime}_{1}M^{\prime}_{2} }=
-D({M^{\prime}_{1}}^{2}+{M^{\prime}_{2}}^{2} )+
J M^{\prime}_{1}M^{\prime}_{2}$ are the unperturbed energy levels.
Taking $S_1=S_2$ in the ensuing discussion,
we calculate
the product (\ref{pro}) and obtain the level splitting
\bena
\Delta E_{-M_1, M_1, M_1,  -M_1}=(4D+2J)
\left({J \over 4D+2J } \right)^{2 M_1} \left[ 
{(S_1 +M_1 )! \over (S_1-M_1)!(2 M_1-1)! }\right]^2.
\label{split1}
\eena
In the ground state ($M_1=S_1$) the result (\ref{split1}) 
simplifies to
\bena
\Delta E_{-S_1, S_1, S_1,  -S_1}=(2S_1)^2 (4D+2J)
\left({J \over 4D+2J } \right)^{2 S_1}.
\label{split1-ground}
\eena
For large value of $M_1$ ($S_1-M_1$, $M_1 \gg 1$) 
Eq. (\ref{split1})  with the help of the Stirling formula reduces to
\bena
\Delta E_{-M_1, M_1, M_1,  -M_1}=\left({2D+J \over \pi} \right)
\left({J \over 4D+2J } \right)^{2 M_1} \left[ 
{(S_1 +M_1 )^{2 S_1+1+M_1} \over (S_1-M_1)^{2S_1+1-M_1} 
(2 M_1)^{4 M_1 -1} }\right].
\eena

Our next example, model II corresponds to the case of
transverse anisotropy in the $xy$-plane:
\bena
{\cal H}^{\rm trans}_{i} =B( \hat{S}^{2}_{xi}-\hat{S}^{2}_{yi}).
\eena
Writing ${\cal H}^{\rm trans}_{i}={1 \over 2} B 
(\hat{S}^{2}_{+i}+\hat{S}^{2}_{-i})$ and choosing
$i=2$, the level splitting 
of the degenerate pair exists
in the matrix elements with connecting the states
$| M^{\prime}_{1}, M^{\prime}_{2}+2k \rangle$
and $| M_{1}, M^{\prime}_{2}+2k+2 \rangle$ where
$k$  is an integer with $0 \leq k \leq (M_2 -M^{\prime}_{2})/2-1$ and
$M^{\prime}_{1}=M_1$.
It corresponds to the level splitting of the degenerate pair
$(M_1,M^{\prime}_{2}) \rightarrow (M_1, M_2)$
where $M_2 > M^{\prime}_{2}$,  $M^{\prime}_{2} <0$,
and $M_2-M^{\prime}_{2}$ is even number.
In the limit $B \ll D$ the level splitting of the degenerate states
appears, minimally, in the $(M_2 -M^{\prime}_{2})/2$-th order in
$B/D$:
\bena
\Delta E_{M^{\prime}_{1} M^{\prime}_{2}, M_1  M_2}&=&
2V_{M^{\prime}_{1} M^{\prime}_{2}, M_{1},M^{\prime}_{2}+2}
{1 \over E_{M_{1}, M^{\prime}_{2}+2}-
E_{M^{\prime}_{1}M^{\prime}_{2} }} 
V_{M_{1}, M^{\prime}_{2}+2, M_{1},M^{\prime}_{2}+4}
\nonumber \\
&&\times 
{1 \over E_{M_{1}, M^{\prime}_{2}+4}-
E_{M^{\prime}_{1}M^{\prime}_{2} }}...
V_{M_{1}, M_{2}-2, M_{1}M_{2}},
\label{pro2}
\eena
where
\bena
V_{M^{\prime}_{1} M^{\prime}_{2}, M_{1},M^{\prime}_{2}+2}&=&
\langle M^{\prime}_{1} M^{\prime}_{2}| 
{B \over 2} \hat{S}^{2}_{-2}
|M_{1},M^{\prime}_{2}+2 \rangle \nonumber \\
&=& {B \over 2} {\tilde{l}}_{M^{\prime}_{2}+1} 
{\tilde{l}}_{M^{\prime}_{2}+2},
\label{v2}
\eena
${\tilde{l}}_{M^{\prime}_{2}}=
\sqrt{ (S_2+M^{\prime}_{2}) (S_2 -M^{\prime}_{2}+1) }$
are the matrix elements of the operator 
$\hat{S}_{-2}$, and
$E_{M^{\prime}_{1}M^{\prime}_{2} }=
-D({M^{\prime}_{1}}^{2}+{M^{\prime}_{2}}^{2} )+
J M^{\prime}_{1}M^{\prime}_{2}-H_z (M^{\prime}_{1}+M^{\prime}_{2})$ 
are the unperturbed energy levels.
Since the pair states are degenerate for the values
of the longitudinal field $H_z=-D(M_2+M^{\prime}_{2})+J M_1$ from
Eq. (\ref{longi-field}), the elements 
$E_{M_1,q}-E_{M^{\prime}_{1}M^{\prime}_{2}}$ in the denomenators of
Eq.  (\ref{pro2})
where $q=M^{\prime}_{2}+2, M^{\prime}_{2}+4...M_2-2$ becomes
independent upon $M^{\prime}_{1}$ and $M_1$. Also,
noting that Eq. (\ref{v2}) is only dependent upon $M^{\prime}_{2}$,
the formula for the level splittings is expected to be
indenpendent of $M^{\prime}_{1}$ and $M_1$ and reads
\bena
\Delta E_{M^{\prime}_{1}, M^{\prime}_2, M_1,  M_2}&=&2D
\left({B \over 2D } \right)^{(M_2-M^{\prime}_{2})/2} 
\sqrt{(S_2+M_2)!(S_2-M^{\prime}_{2})! \over (S_2-M_2)!(S_2+M^{\prime}_{2})! }
\nonumber \\
&& \times 
{ \delta_{M^{\prime}_{1}, M_1} \over 
\left[ (M_2-M^{\prime}_{2} -2)!! \right]^2 },
\label{split2}
\eena
which seems to be the same expression as that
in single-molecule magnet\cite{gar02}.
However, 
the exchange interaction between two single-molecule magnets 
contributes to the 
level splittings via the longitudinal field (\ref{longi-field})
and $\delta_{M^{\prime}_{1}, M_{1}}$ in Eq. (\ref{split2}).

Our final example, model III is described by
\bena
{\cal H}^{\rm trans}=-H_{x} ({\hat S}_{x1}+{\hat S}_{x2}),
\eena
where $H_{x}$ can be internal or external magnetic field.
Using ${\hat S}_{xi}=({\hat S}_{+i}+{\hat S}_{-i})/2$ and
considering the case at $i=2$, 
the level splitting 
of the degenerate pair appears
in the matrix elements with connecting the states
$| M^{\prime}_{1}, M^{\prime}_{2}+k \rangle$
and $| M_{1}, M^{\prime}_{2}+k+1 \rangle$ where
$k$  is an integer with $0 \leq k \leq M_2 -M^{\prime}_{2}-1$ and
$M^{\prime}_{1}=M_1$.
It corresponds to the level splitting of the degenerate pair
$(M_1,M^{\prime}_{2}) \rightarrow (M_1, M_2)$
where $M_2 > M^{\prime}_{2}$, $M^{\prime}_{2} <0$ and
$M^{\prime}_{2}-M_2$ can be any integer.
Thus, the level splitting is represented as
\bena
\Delta E_{M^{\prime}_{1} M^{\prime}_{2}, M_1  M_2}&=&
2V_{M^{\prime}_{1} M^{\prime}_{2}, M_{1},M^{\prime}_{2}+1}
{1 \over E_{M_{1}, M^{\prime}_{2}+1}-
E_{M^{\prime}_{1}M^{\prime}_{2} }} 
V_{M_{1}, M^{\prime}_{2}+1, M_{1},M^{\prime}_{2}+2}
\nonumber \\
&&\times 
{1 \over E_{M_{1}, M^{\prime}_{2}+2}-
E_{M^{\prime}_{1}M^{\prime}_{2} }}...
V_{M_{1}, M_{2}-1, M_{1}M_{2}},
\label{pro3}
\eena
where
\bena
V_{M^{\prime}_{1} M^{\prime}_{2}, M_{1},M^{\prime}_{2}+1}&=&
\langle M^{\prime}_{1} M^{\prime}_{2}| 
-{H_x \over 2} \hat{S}_{-2}
|M_{1},M^{\prime}_{2}+1 \rangle \nonumber \\
&=& -{H_x \over 2} {\tilde{l}}_{M^{\prime}_{2}+1}.
\label{v3}
\eena
Since the unperturbed energy levels $E_{M^{\prime}_{1}M^{\prime}_{2} }$
and the resonant field $H_z$
are the same as the ones in model II,
it is also expected that the level splitting becomes
independent of $M^{\prime}_{1}$ and $M_1$. Therefore,
in the limit of small transverse field  the level splitting is
given by
\bena
\Delta E_{M^{\prime}_{1}, M^{\prime}_2, M_1,  M_2}&=&2D
\left({H_x \over 2D } \right)^{M_2-M^{\prime}_{2}} 
\sqrt{(S_2+M_2)!(S_2-M^{\prime}_{2})! \over (S_2-M_2)!(S_2+M^{\prime}_{2})! }
\nonumber \\
&& \times 
{\delta_{M^{\prime}_{1}, M_1} \over 
\left[ (M_2-M^{\prime}_{2} -1)! \right]^2 },
\label{split3}
\eena
which is similar to that in single-molecule magnet.
This result shows that 
exchange interaction between two single magnets
makes a contribution to the level splitting through the resonant
field, $H_z=-D (M_2+M^{\prime}_{2})+JM_1$ 
and $\delta_{M^{\prime}_{1}, M_{1}}$ in Eq. (\ref{split3}).

Even though we have separately considered the problems
in models II and III, 
both the transverse field and the transverse anisotropy are present in some
cases.  
In the presence of $B$ and
$H_x$ being of the same order of magnitude,
the effect of the transverse field on level splitting is weaker than
that of the transverse anisotropy,
as is evident in Eqs. (\ref{split2}) and (\ref{split3}). 
Thus, we can neglect the transverse field contribution to the splittings.
However, as $M_2-M^{\prime}_{2}$
is odd, the transverse field should be included in the level splitting
through the single perturbation step along the chain
connecting the degenerate states $(M^{\prime}_{1}, M^{\prime}_{2})$
and $(M_1, M_2)$ where $M^{\prime}_{1}=M_1$ and
$M^{\prime}_{2} <0$. 
Hence, the corresponding level splitting becomes
\bena
\Delta E_{M^{\prime}_{1} M^{\prime}_{2}, M_1,  M_2}&=&
2 V^{(H)}_{M^{\prime}_{1} M^{\prime}_{2}, M_{1},M^{\prime}_{2}+1}
{1 \over E_{M_{1}, M^{\prime}_{2}+1}-
E_{M^{\prime}_{1}M^{\prime}_{2} }} 
V^{(B)}_{M_{1}, M^{\prime}_{2}+1, M_{1},M^{\prime}_{2}+3}
...V^{(B)}_{M_{1}, M_{2}-2, M_{1}M_{2}}
\nonumber \\
&+& 
2 V^{(B)}_{M^{\prime}_{1} M^{\prime}_{2}, M_{1},M^{\prime}_{2}+2}
...V^{(B)}_{M_{1}, M_{2}-3, M_{1}M_{2}-1}
{1 \over E_{M_{1}, M_{2}-1}-
E_{M^{\prime}_{1}M^{\prime}_{2} }}
V^{(H)}_{M_{1}, M_{2}-1, M_{1}M_{2}}
\nonumber \\
&+&
2 \sum^{M_2 -3}_{k=M^{\prime}_{2}+2} 
\left( \prod^{k}_{p_1=M^{\prime}_{2}+2} 
V^{(B)}_{M_{1}, p_1-2, M_{1},p_1} \right)
V^{(H)}_{M_{1}, k, M_{1},k+1}
\left( \prod^{M_2-2}_{p_2=k+1} 
V^{(B)}_{M_{1}, p_2, M_{1},p_2+2} \right)
\nonumber \\
&& \times
\left( \prod^{k}_{q_1=M^{\prime}_{2}+2} 
{1 \over E_{M_{1}, q_1}-
E_{M^{\prime}_{1}M^{\prime}_{2} }}  \right)
\left( \prod^{M_2-2}_{q_2=k+1}
{1 \over E_{M_{1}, q_2}-
E_{M^{\prime}_{1}M^{\prime}_{2} }} \right),
\label{pro4}
\eena
where the matrix elements $V^{(B)}$ and $V^{(H)}$ are
expressed as Eqs. (\ref{v2}) and (\ref{v3}), respectively.
The sum in Eq. (\ref{pro4}) can be calculated by
using the formula 
\bena
\sum^{r}_{p=0}{ (2p-1)!! (2r-2p-1)!! \over (2p)!! (2r-2p)!! }=1,
\eena
and the resulting splitting for the odd resonance becomes
\bena
\Delta E_{M^{\prime}_{1}, M^{\prime}_2, M_1,  M_2}&=& H_x
\left({B \over 2D } \right)^{(M_2-M^{\prime}_{2})/2} 
\sqrt{(S_2+M_2)!(S_2-M^{\prime}_{2})! \over (S_2-M_2)!(S_2+M^{\prime}_{2})! }
\nonumber \\
&& \times 
{ \delta_{M^{\prime}_{1}, M_1} \over 
\left[ (M_2-M^{\prime}_{2} -2)!! \right]^2 }.
\label{split4}
\eena

To illustrate the results with concrete example, let us consider
a supramolecular dimer ${\rm Mn_4 O_3 Cl_4 (O_2 CEt)_3 (py)_3}$ 
(hereafter ${\rm Mn_4}$). This compound contains three
${\rm Mn^{3+}}$ ions and one ${\rm Mn^{4+}}$ ion with
the axial anisotropy constant ($D \simeq 0.72$ K), and exchange
coupling ($J \simeq 0.1$ K) between them leads to
the $[{\rm Mn_4}]_{2}$ dimer having a ground
state spin of $S_1=S_2=9/2$. 
At very low temperature, most of the excited states can be
neglected. Thus, as is listed in Table \ref{tab-level}, the low-lying
states are involved in the magnetization reversal at very
low temperature in the presence of the longitudinal field. At high
negative field, the initial state becomes $(-9/2, -9/2)$. As the magnetic
field increases, the first level crossing occurs in the degenerate pair
$(-9/2, -9/2)$, $(-9/2, 9/2)$ at $H_z=-0.336$ T which corresponds
to the odd resonance, i.e., $M_2-M^{\prime}_{2}=$ odd. In this
case the main sources of the level splitting can be either
the transverse anisotropy or the transverse field.
Meanwhile, since the hystersis loops in experiment\cite{wer} display
step-like features at even resonance, $H_z=0.202$ T and 0.873 T,
at least the transverse anisotropy should contribute to  the level splittings.
In this respect, both the transverse anisotropy and the transverse field
induce the level splittings in the odd
resonance and thereby the transition between $(-9/2, -9/2)$ and $(-9/2, 9/2)$.
At the next level crossing at $H_z=0$ T, the degenerate pair is
$(-9/2, -9/2)$ and $(9/2, 9/2)$. 
The possibility of tunneling from $(-9/2, -9/2)$ to $(9/2, 9/2)$  requires
the terms like
either $\hat{S}_{1+}\hat{S}_{2+}$ or $\hat{S}_{1-}\hat{S}_{2-}$ in
the spin Hamiltonian.
However,  there is no such transverse terms
in the Hamiltonian (\ref{hamil}) which  induces the level splitting between
them. For this reason step-like feature is absent in the hystersis loop
at 0 T while a strong quantum step at $H_z=0$ is present in other single-molecule
magnets\cite{tho,san}. 
The situation is analogous to that in the case
$(-9/2, -9/2) \rightarrow (9/2, 7/2)$.
The next level crossing occurs in the degenerate pair
$(-9/2, -9/2)$ and $(-9/2, 7/2)$ which corresponds to the number 2 in Table
\ref{tab-level}. In this situation the level splitting is induced
only by the transverse anisotropy due to the even resonance.
The avoided level crossing at $H_z=0.261$ T not claimed
in the experiment occurs from $(-9/2, 7/2)$ to $(9/2, 7/2)$.
Actually, the corresponding peak position
is shown in the experiment results (Fig. 4 in Ref. \cite{wer}).
The level crossings at $H_z=0.336$ T and 0.739 T allow
tunneling from $(-9/2, 9/2)$ to $(9/2, 9/2)$ and
from $(-9/2, -9/2)$ to $(-9/2, 5/2)$, respectively, which
correspond to the odd resonance.
At $H_z=0.873$ T the avoided level crossing can
occur from $(-9/2, 9/2)$ to $(7/2, 9/2)$ with finite level
splitting which is originated from the transverse anisotropy.
Finally, it is interesting to estimate the level splitting induced by the
transverse exchange interaction. Inserting the value of $D$ and $J$ into
Eq. (\ref{split1-ground}), the level spitting is of the order of $10^{-11}$ K
which is much smaller than that induced by the transverse anisotropy or
the transverse field.
As a result, the main sources of the step-like features in the hystersis
loops of the $[{\rm Mn_4}]_{2}$ dimer are the transverse anisotropy and
the transverse field, and each half of the dimer acts as a field bias
on its neighbor via 
the exchange interaction within
$[{\rm Mn_4}]_{2}$.

In conclusion, we have considered the level splitting
in a dimer with the antiferromagnetic coupling between two
single-molecule magnets. Perturbation approach allows
us to obtain the level splitting of the states degenerate pairwise
for arbitrary spin in the presence of the exchange interaction.
It is found that the level splittings are strongly affected
by the exchange interaction as well as 
the transverse anisotropy and the transverse field.
In comparison with recent experimental results, the level splitting
of the low-lying degenerate pair has been estimated for
several cases and the main sources of each resonance have
been clarified.

This work was supported by grant No. R01-1999-000-00026-0
from the Basic Research Program of the Korea Science and
Engineering Foundation.

\begin{table}
\caption{
The level splitting ($\Delta E$) of the low-lying degenerate pair
$(M^{\prime}_{1}, M^{\prime}_{2})$, $(M_1, M_2)$ in
${\rm Mn_4}$, the resonant 
field ($H_z$) from Eq. (\protect\ref{longi-field}), and
the physical origins which induce  level splittings. 
$B \sim H_x \sim 0.1$ K for illustration.
The numbers, labelled 1 to 5 in the first column
indicate the transitions claimed as the strongest tunnel resonances
in Ref. \protect\cite{wer}. Note that $(M_{1}, M_{2})$
and $(M_{2}, M_{1})$ are degenerate.
}
\label{tab-level}
\end{table}

\begin{center}
\begin{tabular}{|c|c|c|c|c|}  \hline \hline
No. & $(M^{\prime}_{1}, M^{\prime}_{2}) \rightarrow (M_1, M_2)$ 
& $H_z$(T) & $\Delta E$(K) & main sources of splittings \\
\hline
1 & $(-9/2, -9/2) \rightarrow (-9/2, 9/2)$ & $ -0.336$ & $2.02\times 10^{-4}$ & $B$, $H_x$ \\
& $(-9/2, -9/2) \rightarrow (9/2, 9/2)$ &  0 & 0 & -  \\
2 & $(-9/2, -9/2) \rightarrow (-9/2, 7/2)$ &  0.202 & $1.76 \times 10^{-3}$ & $B$  \\
& $(-9/2, -9/2) \rightarrow (9/2, 7/2)$ &  0.233 & 0 & -  \\
& $(-9/2, 7/2) \rightarrow (9/2, 7/2)$ &  0.261 & $2.02\times 10^{-4}$ & $B$, $H_x$   \\
3 & $(-9/2, 9/2) \rightarrow (9/2, 9/2)$ &  0.336 & $2.02\times 10^{-4}$ & $B$, $H_x$  \\
4 & $(-9/2, -9/2) \rightarrow (-9/2, 5/2)$ &  0.739 & $1.19 \times 10^{-3}$ & $B$, $H_x$ \\
5 & $(-9/2, 9/2) \rightarrow (7/2, 9/2)$ &  0.873 & $1.76 \times 10^{-3}$ & $B$   \\
\hline \hline
\end{tabular}
\end{center}

\end{document}